\documentclass[aps,twocolumn,noeprint,superscriptaddress,amsmath,amssymb,showpacs,nolongbibliography]{revtex4-2}
\usepackage{graphicx}
\usepackage{amsmath}
\usepackage{graphics}

\usepackage{amsfonts}
\usepackage{amssymb}
\usepackage{xcolor}
\usepackage{float}

\makeatletter
\newcommand\emailx[1]{%
\move@AF%
\def\@affil{{\normalfont\,#1\strut}{}}%
}%

\usepackage[caption=false]{subfig}
\newcommand{\phantomsubfloat}[1]{
    {
        \captionsetup[subfigure]{labelformat=empty}
        \subfloat[][]{#1}
    }%
}

\usepackage[bookmarks=false]{hyperref}
\hypersetup{colorlinks=true, citecolor=blue, urlcolor=blue, linkcolor=blue}

\newcommand{\pd}{\phantom{\dagger}}
\newcommand{\figref}[1]{(\ref{#1})}

\usepackage[normalem]{ulem}

\begin{document}

\title{Negative Electronic Friction and Non-Markovianity in Nonequilibrium Quantum Systems}

\author{R. J. Preston$^{\dag}$}
\affiliation{Institute of Physics, University of Freiburg, Hermann-Herder-Str.~3, D-79104 Freiburg, Germany}
\author{S. L. Rudge$^{* \: \dag }$}
\affiliation{Institute of Physics, University of Freiburg, Hermann-Herder-Str.~3, D-79104 Freiburg, Germany}
\author{D. S. Kosov}
\affiliation{College of Science and Engineering, James Cook University, Townsville, QLD~4811, Australia \\
\textnormal{${}^{*}$Email: samuel.rudge@physik.uni-freiburg.de} \\
${}^{\dag}$R.J.P and S.L.R contributed equally to this work
}
\author{M. Thoss}
\affiliation{Institute of Physics, University of Freiburg, Hermann-Herder-Str.~3, D-79104 Freiburg, Germany}

\begin{abstract}
\noindent We address the connection between negative electronic friction and non-Markovian effects in the nonadiabatic vibrational dynamics of molecules interacting with metal surfaces under nonequilibrium conditions. We show that a generic nonequilibrium mechanism leading to negative Markovian electronic friction, where molecular vibrations couple directly to inelastic electronic transitions, also introduces significant non-Markovian contributions to the electronic friction. To demonstrate these ideas, we investigate nonequilibrium charge transport through a molecular nanojunction containing a vibrationally coupled donor-acceptor model, where negative electronic friction reflects driving of the vibrational mode beyond standard Joule heating. By comparison to numerically exact, fully quantum hierarchical equations of motion simulations, we verify that these non-Markovian effects have a significant impact on the nonequilibrium dynamics and even on the stability of the resulting Langevin equation.
\end{abstract}

\maketitle

\noindent The dynamics of nuclear degrees of freedom coupled to extended electronic states is often highly nonadiabatic in nanoscale systems, as dense electronic spectra and nonequilibrium electronic excitations can induce transitions between adiabatic potential energy surfaces even for slow nuclear motion. Understanding and modeling these processes is thus critical in a wide variety of scenarios \cite{Hou2025_unraveling,Li2025_exciton-phonon,Klinduhov2016_vibronic,Wang2025_manipulating,Frezza2025}, and is of particular importance for nonequilibrium transport in molecular nanojunctions, where current-driven electronic excitations strongly influence vibrational relaxation, stability, and energy flow during charge transport, necessitating theoretical descriptions beyond the Born–Oppenheimer approximation.

The electronic friction and Langevin dynamics (EFLD) approach is one of the most popular mixed quantum-classical (MQC) methods used to simulate the nonadiabatic dynamics out of equilibrium. Such MQC methods are obtained by treating molecular vibrations classically and integrating out quantum electronic degrees of freedom (DoFs). In the EFLD approach, a further limit of weak nonadiabaticity is enforced, yielding a stochastic Langevin equation for the vibrational DoFs in which the quantum electronic DoFs appear as effective electronic forces \cite{Dou2017a,Rudge2023,Lue2012,Trenins_nonmarkovian_2025,Chen2019a,Chen2019b}:
\begin{align}
    m \ddot{x} = -\partial_{x}U_{\text{vib}} + F^{\text{ad}}_{\text{el}}(x) - \int^{t}_{0} d\tau \: \gamma(x,t-\tau) \dot{x}(\tau) + f(t), \label{eq: non-Markovian Langevin quasistationary limit}
\end{align}
where we have included a single classical vibrational DoF with mass $m$ and conjugate position and momentum $(x,p)$ for simplicity. 

Although there are multiple approaches to enforcing weak nonadiabaticity, Eq.\eqref{eq: non-Markovian Langevin quasistationary limit} has been derived via a timescale separation between fast electronic and slow vibrational DoFs, specifically in the quasi-stationary limit where two-time functions become functions of time differences only. Consequently, $-\partial_{x}U_{\text{vib}}$ and $F^{\text{ad}}_{\text{el}}(x)$ represent the unperturbed vibrational force and the adiabatic contribution to the mean electronic force, respectively. Meanwhile, nonadiabatic effects are included through the electronic friction (EF) coefficient, $\gamma(x,t-\tau)$, and the zero-mean Gaussian stochastic force, $f(t)$, with correlation function $\langle f(t)f(\tau) \rangle = D(x(t),t-\tau)$. Here, $x(t)$ refers to the current vibrational coordinate.

In most practical applications, Eq.\eqref{eq: non-Markovian Langevin quasistationary limit} is only solved in the Markovian limit \cite{HeadGordon1995,Maurer2016,Rudge2023,Rudge2024,Dou2015d,Dou2016a,Maeck2024},
\begin{align}
    m\ddot{x} = -\partial_{x}U_{\text{vib}} + F^{\text{ad}}_{\text{el}}(x) - \tilde{\gamma}(x,0) \dot x + f(t), \label{eq: Markovian Langevin equation}
\end{align}
where $\tilde{\gamma}(x,0)$ represents the zero-frequency component of the EF spectrum,
\begin{align}
 \tilde \gamma (x,\omega) = \int d\tau e^{i\omega \tau} \gamma(x,\tau),   
\end{align}
and the Gaussian stochastic force is sampled in the same limit: $\langle f(t) f(\tau) \rangle = \tilde{D}(x,0) \delta(t-\tau)$. Details of these limits and explicit expressions for all quantities are shown in the Supplementary Material. 

Even in the Markovian limit, electronic friction contains a significant amount of physical information, as it represents a first-order nonadiabatic correction to $F^{\text{ad}}_{\text{el}}(x)$. For example, in equilibrium, $\tilde{\gamma}(x,0)$ captures the dissipative effect of vibrational energy loss owing to electron-hole pair (EHP) excitation \cite{HeadGordon1995}, which is balanced by the influence of the stochastic force via the fluctuation-dissipation theorem (FDT). In contrast, nonequilibrium dynamics can break the FDT, which can, for example, lead to noise-driven Joule heating of the molecular vibrations \cite{Chen2019b,Dou2017c,Rudge2024,Preston2023}. Furthermore, $\tilde{\gamma}(x,0)$ is \textit{not} guaranteed to be positive-definite out of equilibrium, leading to the possibility of exotic sources of vibrational heating in molecular nanojunctions \cite{Maeck_Vibrational2025,Lue2010,Lue2011,Lue2012,Lue2019,Bode2011,Bode2012,Todorov2011,Todorov2014, Preston2020, Preston2022}. 

Such a generic mechanism leading to negative Markovian EF forms the focus of this work, in which the vibrational mode couples directly to inelastic electronic transitions \cite{Lue2011,Bode2011,Bode2012}. Under nonequilibrium conditions, these transitions are biased, and energy is thus pumped deterministically into the vibrational mode via EHP creation. Because the vibrational coordinate directly modulates the transition itself, this energy transfer enters the mean electronic force, giving rise to a negative Markovian EF coefficient rather than mere stochastic heating.

We show, however, that the same inelastic electronic transitions that give rise to negative Markovian EF also introduce additional electronic relaxation timescales, which can in turn introduce significant non-Markovian effects into the electronic forces. Via a basic model of nonequilibrium charge transport through a vibrationally coupled molecular bridge \cite{Simine2012,simine_vibrational_2012,segal_heat_2005,segal_heat_2006,segal_nonequilibrium_2011,Foti2018,foti_interface_2017}, we demonstrate that these non-Markovian effects can actually dominate the Markovian contribution to the electronic forces, which has deep implications for the physical interpretation of negative EF and the stability of the resulting Langevin equation. 

The physical ingredients underlying the non-Markovianity and negative EF presented here are expected to arise broadly in nonequilibrium systems that bias vibrationally coupled inelastic electronic transitions, such as in more complicated molecular bridges, nanojunctions with position-dependent molecule-lead couplings \cite{Preston2022}, light-driven processes at metal surfaces \cite{Lue2011}, and interacting systems with multiple electron-electron \cite{Dou2017c,Rudge2023,Rudge2024} or electronic-vibrational \cite{Rudge2023,Rudge2024,Maeck2024} quantum timescales. 


The general molecular nanojunction Hamiltonian is
\begin{align}
H = \: & H_{\text{mol}} + H_{\text{leads}} + H_{\text{mol-leads}},
\end{align}
where $H_{\text{mol}}$ is the Hamiltonian of the molecule, $H_{\text{leads}}$ is the Hamiltonian of the leads, and $H_{\text{mol-leads}}$ is the interaction between them. In this work, the molecular Hamiltonian describes a vibrationally coupled donor-acceptor model,
\begin{align}
H_{\text{mol}} = \: & \Delta \left(d^{\dag}_{1}d^{\pd}_{1} - d^{\dag}_{2}d^{\pd}_{2}\right) + \lambda \sqrt{2} \left(d^{\dag}_{1}d^{\pd}_{2} + d^{\dag}_{2}d^{\pd}_{1}\right)\hat{x}  
\nonumber \\
& + \frac{\Omega}{2}\left(\hat{x}^{2} + \hat{p}^{2}\right), \label{eq: molecular Hamiltonian}
\end{align}
where the operators $d^{\dag}_{m}$ and $d^{\pd}_{m}$ create and annihilate an electron in the $m$th electronic state with energy $\varepsilon^{\pd}_{m}$, respectively, while $\{\hat{x},\hat{p}\}$ are the dimensionless position and conjugate momentum of the harmonic vibrational mode with frequency $\Omega$. This mode is coupled linearly to the hopping between the electronic levels, with coupling strength $\lambda$. The leads are modeled as reservoirs of noninteracting electrons with Hamiltonian 
\begin{align}
H_{\text{leads}} = \: & \sum_{\alpha \in \{L,R\} }\sum_{k} \varepsilon^{\pd}_{k} c^{\dag}_{k\alpha}c^{\pd}_{k\alpha},
\end{align}
where $c^{\dag}_{k\alpha}$ and $c^{\pd}_{k\alpha}$ create and annihilate an electron with energy $\varepsilon_{k}$ in lead $\alpha$. The leads are assumed to be held at local equilibrium with temperature $T$ and chemical potentials $\mu_{\alpha}$. Nonequilibrium transport conditions are enforced via a voltage bias across the junction, $\Phi = (\mu_{L} - \mu_{R})/e$, which is induced by varying the chemical potentials symmetrically around zero: $\mu_{L} = e\Phi/2 = -\mu_{R}$. The interaction between these two subsystems is governed by the molecule-lead interaction term,
\begin{align}
H_{\text{mol-leads}} = \: & \sum_{k} \left[V^{\pd}_{kL,1}\left(c^{\dag}_{kL}d^{\pd}_{1} + d^{\dag}_{1}c^{\pd}_{kL}\right) + \right. \nonumber \\
& \qquad \left. V^{\pd}_{kR,2}\left(c^{\dag}_{kR}d^{\pd}_{2} + d^{\dag}_{2}c^{\pd}_{kR}\right)\right].
\end{align}
Here, $V^{\pd}_{k\alpha,m}$ represents the coupling strength between state $k$ in lead $\alpha$ and state $m$ in the molecule. 

\begin{figure*}
    \begin{center}
       \phantomsubfloat{\label{fig: vib energy schematic a}}
       \phantomsubfloat{\label{fig: vib energy schematic b}}
       \phantomsubfloat{\label{fig: vib energy schematic c}}
       \includegraphics[width=\textwidth]{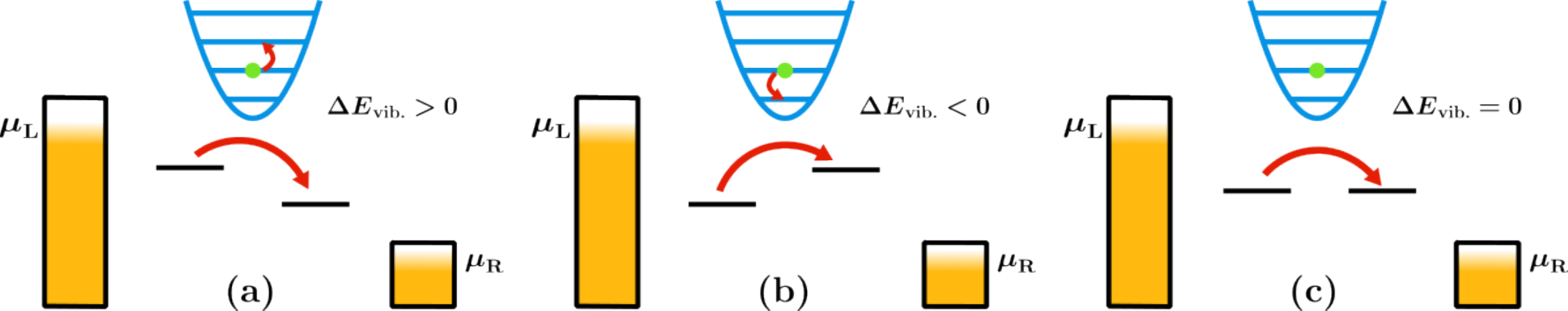}
       \caption{Three different energetic configurations of the donor-acceptor model: (a) $\Delta > 0$, (b) $\Delta <0$, and (c) $\Delta = 0$.}
       \label{fig: vib energy schematic}
    \end{center}
\end{figure*}

A schematic of this model is shown in Fig.~\figref{fig: vib energy schematic} for three different electronic configurations. Given that the only pathway for electron transport is via the vibrationally mediated hopping term, the transport is firmly in the sequential tunneling regime for weak electron-vibrational coupling, $\lambda < \Gamma$. At finite bias and for $\Delta > 0$, as depicted in Fig.~\figref{fig: vib energy schematic a}, electrons must therefore inelastically transport through the system, which constantly pumps energy into the vibration. Conversely, for $\Delta < 0$, shown in Fig.~\figref{fig: vib energy schematic b}, electron transport is facilitated by the dissipation of energy from the vibrational mode. Finally, in Fig.~\figref{fig: vib energy schematic c}, the donor and acceptor states are equal in energy, so that electron transport applies only a Joule heating effect to the molecular vibration. 

\begin{figure}
    \begin{center}
    \includegraphics[width=0.99\linewidth]{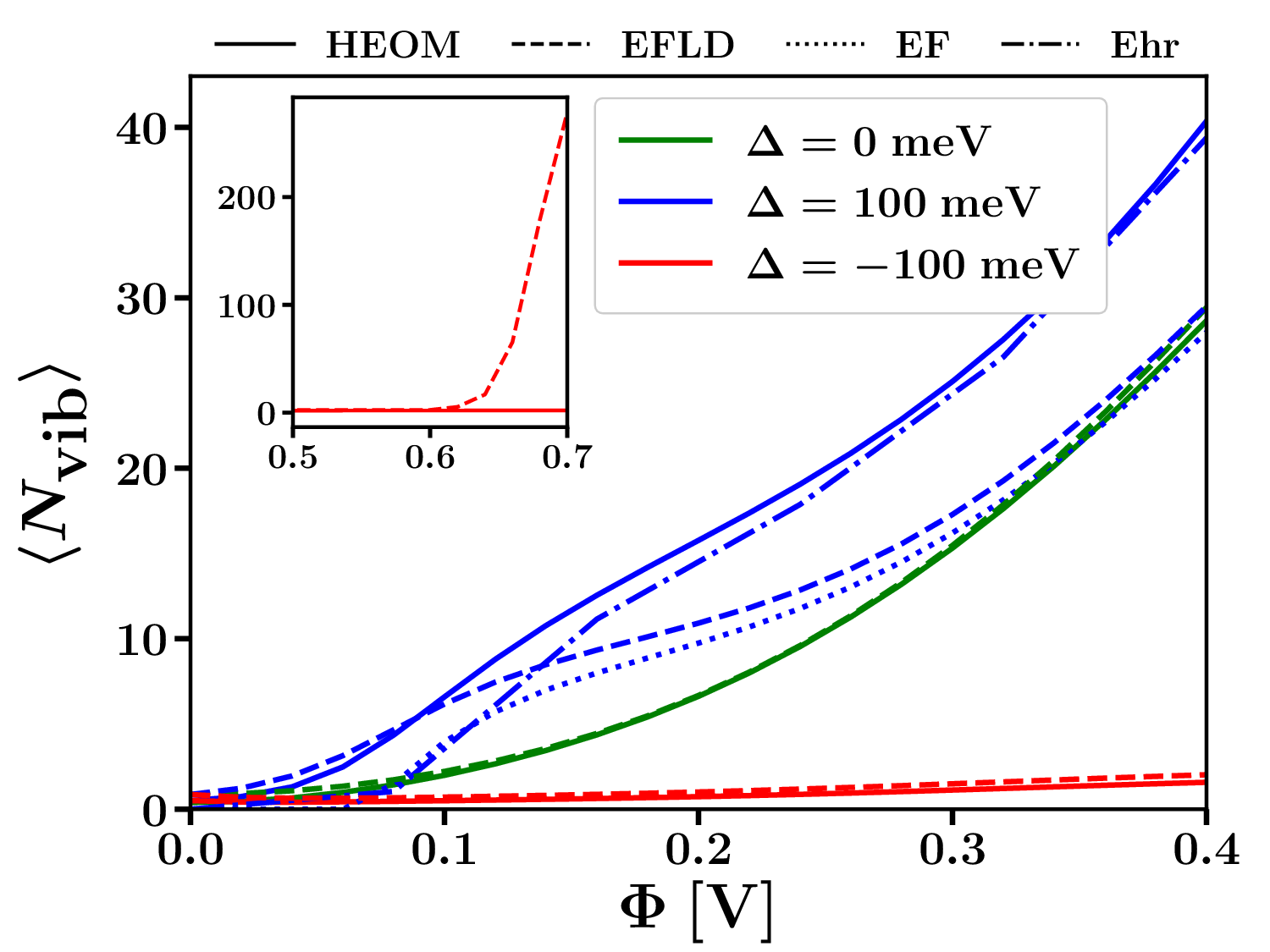}
    \caption{Steady-state average vibrational excitation, $\langle N_{\text{vib}} \rangle$, as a function of bias voltage. Inset: $\langle N_{\text{vib}} \rangle$ at higher voltages for $\Delta = -100\text{ meV}$.  Parameters: $\Gamma = 100$ meV, $\lambda = 10$ meV, $\Omega = 30$ meV, and $k_{B}T = 25.8 \text{ meV}$. The solid lines are calculated fully quantum mechanically with HEOM, while the dashed, dotted, and dash-dot lines refer to results obtained from Markovian electronic friction and Langevin dynamics (EFLD), Markovian electronic friction without the stochastic force (EF), and the Ehrenfest approach (Ehr), respectively.}
    \label{fig: quantum results}
    \end{center}
\end{figure}

This is demonstrated in Fig.~\figref{fig: quantum results}, in which the nonequilibrium steady-state average vibrational excitation, $\langle N_{\text{vib}} \rangle$, is plotted against the bias voltage. We first focus on the quantum results (solid lines) which are obtained from numerically exact hierarchical equations of motion (HEOM) simulations, with details given the Supplementary Material. In dimensionless coordinates, the quantum vibrational excitation operator is defined as {$\hat{N}_{\text{vib}} = \hat{E}_{\text{vib}} / \Omega = \frac{1}{2}\left(\hat{x}^{2} + \hat{p}^{2} \right)$} and the corresponding expectation value is calculated via the steady-state molecular density matrix as $\langle N_{\text{vib}} \rangle = \text{Tr}_{\text{mol}} \left\{\hat{N}_{\text{vib}} \rho^{\text{ss}}_{\text{mol}}\right\}$. Matching the energy schematic in Fig.~\figref{fig: vib energy schematic}, we observe that $\langle N_{\text{vib}} \rangle$ is strongly enhanced (suppressed) in the $\Delta = 100\text{ meV}$ ($\Delta = -100\text{ meV}$) case when compared to the $\Delta = 0$ case, which contains only Joule heating. 

We now investigate whether such vibrational heating and cooling effects are captured within the Markovian EFLD approach of Eq.\eqref{eq: Markovian Langevin equation} (dashed lines), with additional comparison to simulations obtained from Ehrenfest dynamics (dash-dot lines). These serve as an important reference point, as the Ehrenfest approach neglects the stochastic force but includes the exact non-Markovian, path-dependent mean electronic force:
\begin{align}
    m\ddot{x} = -\partial_{x}U_{\text{vib}} + \langle F^{\pd}_{\text{el}} \rangle [x(t)].
\end{align}
In the MQC simulations, the classical vibrational excitation operator is obtained by replacing $\left\{\hat{x},\hat{p}\right\} \rightarrow \left\{x,p\right\}$ in $\hat{N}_{\text{vib}}$, and $\langle N_{\text{vib}} \rangle$ is then simulated via an ensemble average over classical trajectories, with details given in the Supplementary Material. 

In the $\Delta > 0$ case (blue lines), Ehrenfest dynamics reproduces the quantum results accurately at large voltages, indicating that quantum contributions from the vibrational mode are minimal and that the dominant vibrational excitation mechanism is indeed the deterministic pumping process demonstrated in Fig.~\figref{fig: vib energy schematic a}. Conversely, at low bias voltage where Joule heating dominates, the Ehrenfest approach underestimates $\langle N_{\text{vib}} \rangle$, as the method explicitly neglects the electronic force fluctuations that drive vibrational heating. This contrasts with EFLD, which reproduces the quantum result well at low bias voltage but underestimates $\langle N_{\text{vib}} \rangle$ at larger bias voltages, where the mechanism of vibrational excitation induced by $\Delta$ dominates. This observation is strengthened by the dotted lines, which were obtained by simulating Eq.\eqref{eq: Markovian Langevin equation} \textit{without} the stochastic force, and which lie almost exactly on top of the results of the full stochastic EFLD. In the following, we will demonstrate that this large vibrational excitation manifests itself as a negative Markovian EF coefficient.

We now turn to the $\Delta < 0$ case, where the mechanism of vibrational cooling induced by the donor-acceptor coupling strongly damps vibrational motion and Joule heating is the only source of vibrational excitation. Since Joule heating is contained within the stochastic force, only the EFLD approach is shown for $\Delta = -100\text{ meV}$, and it reproduces the quantum result well for small voltages. However, as shown in the inset, after some critical voltage has been reached, EFLD becomes unstable, generating extremely large $\langle N_{\text{vib}}\rangle$ and rapidly diverging from the exact quantum result. In what follows, we will also show that this behavior is a manifestation of negative EF, but that it is \textit{unphysical} and suggests that the Markovian EFLD approach breaks down, as these high excitations do not appear in the quantum simulations. Finally, when $\Delta = 0$, Joule heating dominates and the EFLD method performs well, quantitatively reproducing the quantum result for all voltages. 

\begin{figure}
    \begin{center}
    \phantomsubfloat{\label{fig: frics_abc a}}
    \phantomsubfloat{\label{fig: frics_abc b}}
    \phantomsubfloat{\label{fig: frics_abc c}}
    \phantomsubfloat{\label{fig: frics_abc d}}
    \includegraphics[width=0.8\linewidth]{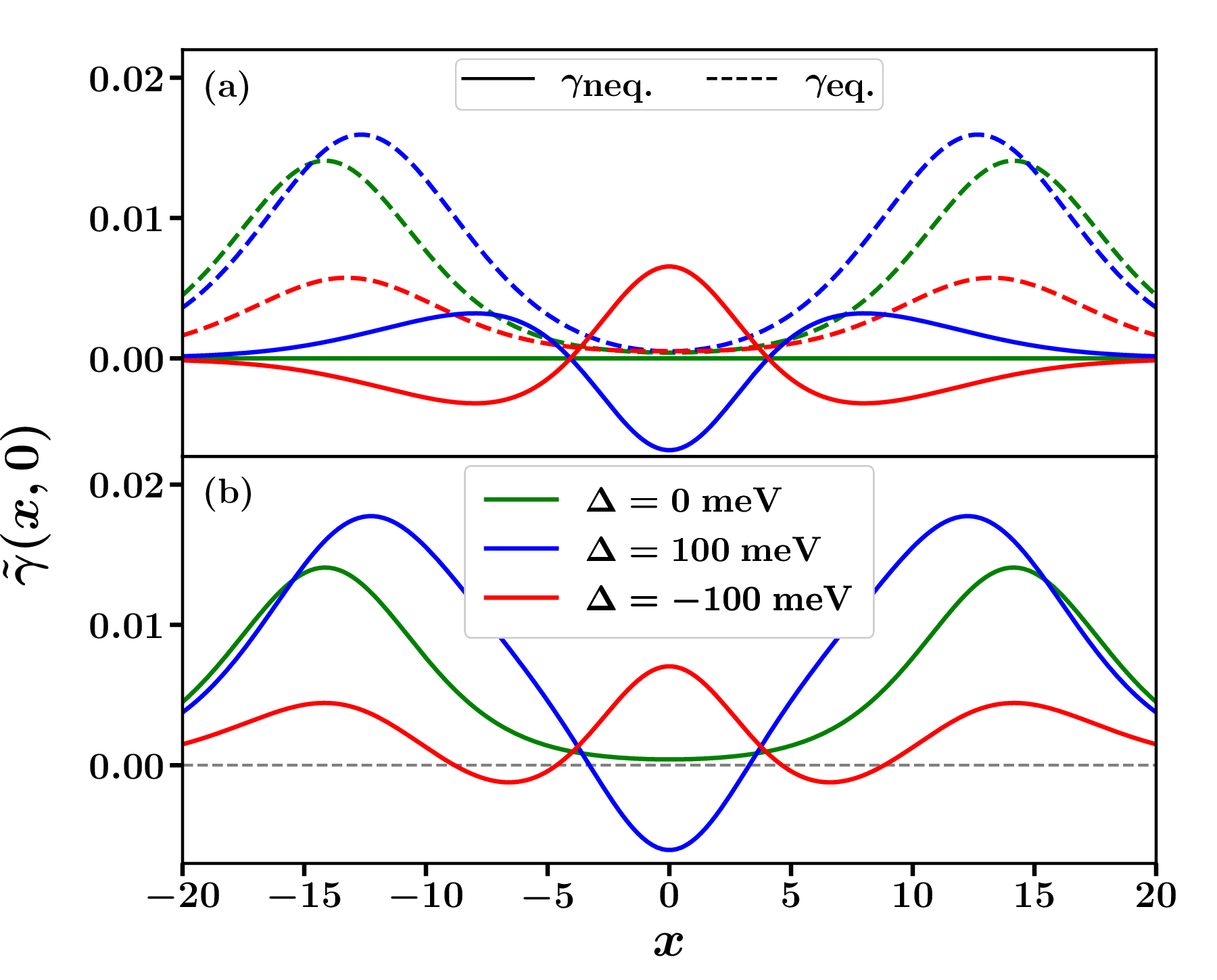}
    \includegraphics[width=0.75\linewidth]{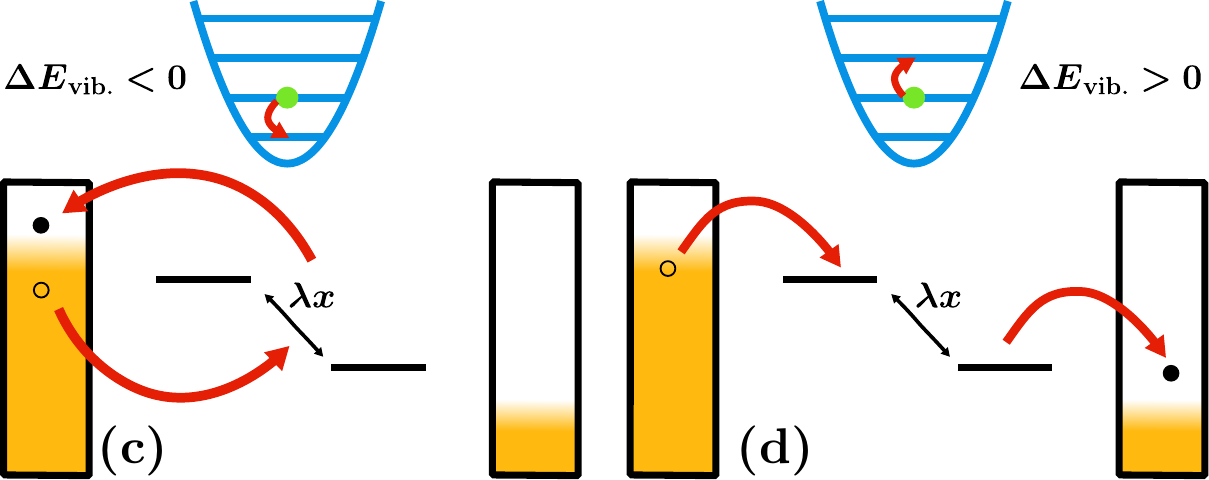}
    \caption{(a) Equilibrium and nonequilibrium contributions of the Markovian friction coefficient as a function of $x$ and (b) total Markovian friction coefficient, computed at $\Phi = 0.4$ V. Other parameters are the same as in Fig.~\figref{fig: quantum results}. Schematics of equilibrium and nonequilibrium EHP creation processes are shown in (c) and (d), respectively.}
    \label{fig: frics_abc}
    \end{center}
\end{figure}

Next, we connect this mechanism of vibrational excitation directly to EHP processes and the Markovian electronic forces, which can be partitioned into equilibrium and nonequilibrium components:
\begin{align}
    \tilde{\gamma}(x,0) = \: &{\gamma}_\text{eq.}(x) + {\gamma}_\text{neq.} (x) \\
    \tilde{D}(x,0) = \: & {D}_\text{eq.}(x) + {D}_\text{neq.} (x).
\end{align}
We specify the equilibrium components as those that satisfy the FDT, ${D}_\text{eq.}(x) = k_{B}T{\gamma}_\text{eq.}(x)$, which holds even in nonequilibrium (shown in Supplementary Material). Vibrational heating in nonequilibrium then arises via two distinct mechanisms in a 1D system. The first is Joule heating, which occurs when $0 \leq k_{B}T {\gamma}_\text{neq.} (x) <{D}_\text{neq.} (x)$. The second is a deterministic heating due to the electronic geometry of the molecular bridge, which occurs when $k_{B}T\gamma_\text{neq.} (x) < 0 \leq D_\text{neq.} (x)$. Notably, ${D}_\text{neq.} (x)$ is proportional to $\Phi^2$ to leading order, whereas ${\gamma}_\text{neq.} (x)$ is proportional to $\Phi$ to leading order, as is demonstrated in the Supplementary Material. This may offer a natural explanation for experimental observations of two distinct mechanisms of bond breakage occuring at different voltages \cite{Sabater2015}. For an explicit demonstration of these ideas and how negative Markovian EF arises, we now examine the equations for ${\gamma}_\text{eq.}(x)$ and ${\gamma}_\text{neq.}(x)$ at $x = 0$, where the equations for arbitrary $x$ are shown in the Supplementary Material:
\begin{align}
   {\gamma}_\text{eq.}(0) = \: \frac{-\lambda^2}{2\pi} 
     \int^\infty_{-\infty}d\epsilon \: A_\text{d}(\epsilon)A_\text{a}(\epsilon) \Big(\frac{\partial f_L}{\partial \epsilon} + \frac{\partial f_R}{\partial \epsilon}\Big), 
    \label{gamma_eq}
\end{align}
Here, we have introduced the spectral functions for the lead-hybridized donor and acceptor states, 
\begin{align}
    A_\text{d/a}(\epsilon) = \Gamma [(\epsilon\mp \Delta)^2 + \Gamma^2 / 4]^{-1},
\end{align}
with $\mp$ for donor and acceptor, respectively, as well as the Fermi-Dirac functions, $f_{\alpha}(\epsilon) = [e^{(\epsilon - \mu_{\alpha})/k_{B}T} + 1]^{-1}$. The form of Eq.\eqref{gamma_eq} shows that ${\gamma}_\text{eq.}(0)$ contains additive contributions from each lead individually, encoding the equilibrium, dissipative contribution of EHP excitation within one lead, shown schematically in Fig.~\figref{fig: frics_abc c}. Note that, even though these are equilibrium processes, they also exist in nonequilibrium and are maximized at vibrational coordinates where the corresponding donor-acceptor eigenenergies coincide with $\mu_{\alpha}$ \cite{Rudge2023,Dou2015d}. This is demonstrated in the dashed lines of Fig.~\figref{fig: frics_abc a}, which shows ${\gamma}_\text{eq.}(x)$ and ${\gamma}_\text{neq.}(x)$ for general $x$ and at $\Phi = 0.4\text{ V}$.

Likewise, the nonequilibrium component at $x = 0$ is given by
\begin{align}
    &{\gamma}_\text{neq.}(0) =  \frac{\lambda^2}{4\pi} \int^\infty_{-\infty}d\epsilon \Big(f_L - f_R\Big) \Big(A_\text{d}\frac{\partial A_\text{a}}{\partial \epsilon} - A_\text{a}\frac{\partial A_\text{d}}{\partial \epsilon} \Big). 
    \label{eq: noneq friction at x = 0}
\end{align}
From the structure of Eq.\eqref{eq: noneq friction at x = 0}, it is evident that the sign of ${\gamma}_\text{neq.}$, and therefore its dissipative or driving nature, is determined both by the voltage polarity and by the sign of $\Delta$. In contrast to ${\gamma}_\text{eq.}$, $\tilde{\gamma}_\text{neq.}$ describes a vibrationally mediated EHP creation process across the junction that induces an inelastic electronic transition within the donor-acceptor model, shown schematically in Fig.~\figref{fig: frics_abc d}. Consequently, at positive bias and $x = 0$, the $\Delta > 0 $ case induces negative ${\gamma}_\text{neq.} (x)$, which is also observable in the solid blue line of Fig.~\figref{fig: frics_abc a} in the region around $x = 0$. Since ${\gamma}_\text{eq.}$ is negligible in this region, the total Markovian EF $\tilde{\gamma}(x,0)$ is also negative here, as is demonstrated in Fig.~\figref{fig: frics_abc b}, which drives the vibrational dynamics before stabilizing to a stationary state where $\tilde{\gamma} > 0$. 

This behavior is inverted for $\Delta < 0$, which is observable from Eq.\eqref{eq: noneq friction at x = 0} as well as the region around $x = 0$ in the solid red line of Fig.~\figref{fig: frics_abc a}. Here, the EHP-mediated process induces an inelastic electronic transition that dissipates energy from the vibrational mode. Notably, however, ${\gamma}_\text{neq.}$ becomes \emph{negative} at larger $x$-values, where the donor and acceptor states are strongly hybridized. These regions of negative Markovian EF are what lead to the breakdown of EFLD observed at sufficiently high voltage in Fig.~\figref{fig: quantum results}. Finally, we note that ${\gamma}_\text{neq.}$ disappears in equilibrium or when $\Delta =0$, as in the solid green line of Fig.~\figref{fig: frics_abc a}. 

Although the mechanism of vibrational excitation (dissipation) due to $\Delta > 0$ ($\Delta < 0$) is \textit{qualitatively} captured within the Markovian EF around $x = 0$, they are clearly not captured with the same fidelity as the quantum simulations, especially at larger $|x|$. A clue to this discrepancy lies in the strong agreement between Ehrenfest dynamics and the quantum result for the $\Delta > 0$ configuration, implying that the discrepancies must be due to either higher-order nonadiabatic corrections or non-Markovian effects, as these are captured in $F_{\text{el}}[x(t)]$. In the following, we argue that non-Markovian contributions arising from the $\Delta^{-1}$ timescale to the EF coefficient are likely to be critical, even though we operate in a regime traditionally regarded as near-adiabatic: $\Gamma = 100\text{ meV} \gg \Omega = 30 \text{ meV}$ \cite{Rudge2024}.


We frame this analysis with the time-averaged power density dissipated by the EF force,
\begin{align}
    \bar{P}_{\text{diss.}}(x,\omega) = \: & |\tilde{\dot{x}}(\omega)|^{2} \text{Re}\left\{\tilde{\gamma}(x,\omega)\right\}, \label{eq: power dissipated spectrum}
\end{align}
which shows that frequency $\omega$ at position $x$ has a dissipative contribution if $\text{Re}\left\{\tilde{\gamma}(x,\omega)\right\} > 0$. The Markovian limit would thus perform well if the vibrational dynamics is dominated by frequencies around $\omega \approx 0$ and if $\text{Re}\{\tilde{\gamma}(x,\omega)\}$ does not show any additional significant structure besides $\tilde{\gamma}(x,0)$.

\begin{figure}
    \begin{center}
       \phantomsubfloat{\label{fig: fric_freq_gam1 a}}
       \phantomsubfloat{\label{fig: fric_freq_gam1 b}}
       \phantomsubfloat{\label{fig: fric_freq_gam1 c}}
    \includegraphics[width=0.99\linewidth]{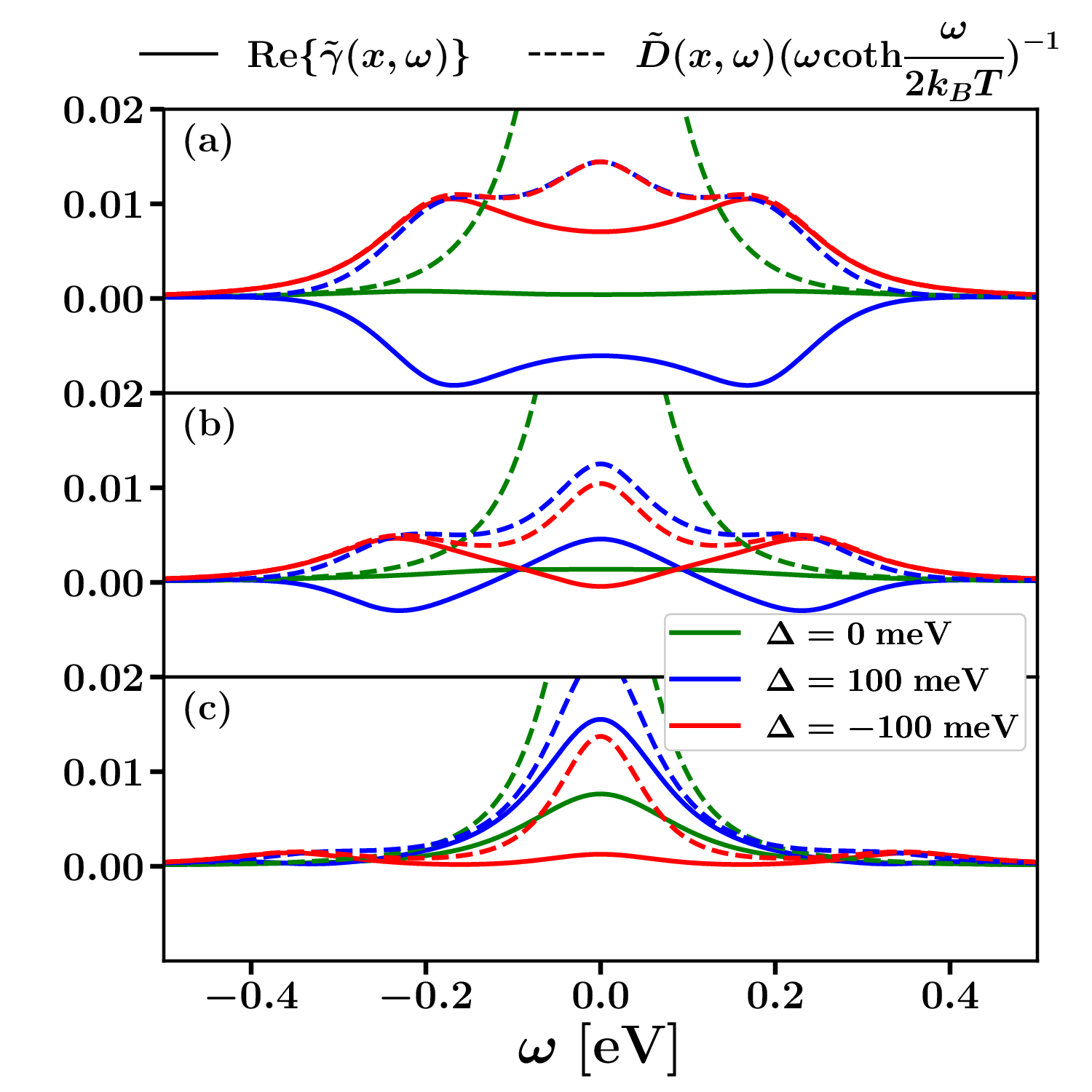}
    \caption{Real part of the Fourier transform of the friction, $\text{Re}\left\{\tilde{\gamma}\right\}$ and corresponding relation to the correlation function of the stochastic force with $\Phi = 0.4$ V at three different vibrational coordinates: (a) $x= 0$, (b) $x = 5$, (c) $x = 10$. Other parameters are the same as in Fig.~\figref{fig: quantum results}.}
    \label{fig: fric_freq_gam1}
    \end{center}
\end{figure}

We explore this assumption in Fig.~\figref{fig: fric_freq_gam1}, which plots $\text{Re} \left\{\tilde{\gamma}(x,\omega) \right\}$ at $\Phi = 0.4\text{ V}$, and for the three energetic configurations of the molecular bridge and at three representative vibrational coordinates: (a) $x = 0$, (b) $x = 5$, and (c) $x = 10$. First, one observes that the Markovian approximation \textit{is} justified for $\Delta = 0$ (green lines), as there is a single peak at $\omega = 0$ that dominates all other frequencies. This matches the results of Fig.~\figref{fig: quantum results}, in which the Markovian electronic friction approach reproduce the quantum simulations well. 

Next, we consider the $\Delta > 0$ configuration (blue lines). While the $\omega = 0$ contribution is indeed negative and significant when $x=0$ in (a), there are two additional peaks that are also negative and larger in magnitude at $\omega \approx \pm 2\Delta = \pm 200\text{ meV}$. Therefore, while the Markovian contribution to the electronic friction has a driving effect, including non-Markovian effects from other frequencies could \textit{increase} the vibrational excitation. Furthermore, as shown in (b) at $x = 5$, these side-peaks can even have a different sign to Markovian contribution. For $\Delta > 0$, this indicates that including non-Markovian effects may increase the coordinate range over which electronic friction drives vibrational excitation. Only at large values of the vibrational coordinate does the Markovian assumption become satisfied, as shown in (c) for $x = 10$. Here, the electronic eigenstates of the donor-acceptor model are near $\mu_{\alpha}$, where EHP excitation dominates.

An inverse result is found for the $\Delta < 0$ configuration (green lines), where non-Markovian effects would effectively shrink the range of vibrational coordinates over which the EF drives vibrational excitation. This would explain the apparent breakdown of EFLD in Fig.~\figref{fig: quantum results}, as it indicates that the negative EF may simply be an artifact of the Markovian approximation, and that including non-Markovian effects could produce an overall \textit{dissipative} effect at coordinates that were previously driven by $\tilde{\gamma}(x,0) < 0$. Finally, we note that this discussion applies equally to the $\tilde{D}(x,\omega)$ which, although always guaranteed to be positive-definite, displays additional structure for $\Delta \neq 0$. The units for $\tilde{D}(x,\omega)$ are chosen to represent the difference to the equilibrium quantum FDT, which at finite frequency takes the form $\tilde{D}(x,\omega) = \: \omega \coth(\frac{\omega}{2 k_{B}T })\text{Re}\left\{\tilde{\gamma}(x,\omega)\right\}$.

In summary, we have shown that whenever a classical degree of freedom couples to inelastic fermionic transitions across an energy gap $\Delta$ in a molecular nanojunction, and nonequilibrium conditions bias excitations against de-excitations, the same mechanism producing negative Markovian EF necessarily generates a non-Markovian spectral structure at frequencies determined by $\Delta$. This structure can dominate the zero-frequency component and even carry the opposite sign, so that the dissipative or driving character of the fermionic forces cannot be inferred from the Markovian EF alone. The ingredients for this process are not specific to the molecular nanojunction studied here but arise generally across nonequilibrium physics, from nonadiabatic molecular dynamics at metal surfaces and current-driven nanoelectromechanical systems to nuclear fission and impurity dynamics in driven Fermi gases. Our results therefore reframe negative EF from a single number characterizing local driving or dissipation to a frequency-resolved spectral landscape whose integrated effect governs the true classical dynamics. Our analysis highlights the complexity of the term “negative electronic friction” and the difficulty in predicting vibrational dynamics from such coarse-grained quantities alone. 


\section*{Supporting Information}

\noindent Expanded theory and numerical details of quantum and mixed quantum-classical simulations.

\section*{Acknowledgements}
This work was supported by the Deutsche Forschungsgemeinschaft (DFG) within the framework of the Research Unit FOR5099 “Reducing complexity of nonequilibrium systems”. R.J.P thanks the Alexander von Humboldt Foundation for the award of a Research Fellowship. The authors acknowledge the support by the state of Baden-Württemberg through bwHPC and the DFG through Grant No. INST 40/575-1 FUGG (JUSTUS 2 cluster).

\appendix


\bibliography{Main_text_incl._fig.bib}

\end{document}